\documentclass[aps, showpacs, groupedaddress, twocolumn]{revtex4}
\usepackage[colorlinks, breaklinks, citecolor=blue, linkcolor=blue, urlcolor={blue}]{hyperref}
\usepackage{amsfonts}
\usepackage{amsmath}
\usepackage{mathrsfs}
\usepackage{amssymb}
\usepackage[dvips]{graphics,color}
\usepackage{graphicx}
\usepackage{dcolumn}
\usepackage{bm}

\begin{document}
	
\title{Nonreciprocal Phonon Laser}

\author{Y. Jiang$^1$, S. Maayani$^2$, T. Carmon$^2$, Franco Nori$^{3,5}$, and H. Jing$^{4}$}

\email[]{jinghui73@foxmail.com}
\affiliation{$^1$Department of Physics, Henan Normal University, Xinxiang 453007, China}
\affiliation{$^2$Mechanical Engineering, Technion - Israel Institute of Technology, Haifa 32000, Israel}
\affiliation{$^3$Theoretical Quantum Physics Laboratory, RIKEN Cluster for Pioneering Research, Wakoshi, Saitama 351-0198, Japan}
\affiliation{$^4$Department of Physics and Synergetic Innovation Center for Quantum  Effects and Applications, Hunan Normal University, Changsha 410081, China}
\affiliation{$^5$Physics Department, The University of Michigan, Ann Arbor, Michigan 48109-1040, USA}

\date{\today}

\begin{abstract}
We propose nonreciprocal phonon lasing in a coupled cavity system composed of an optomechanical and a spinning resonator. We show that the optical Sagnac effect leads to significant modifications in both the mechanical gain and the power threshold for phonon lasing. More importantly, the phonon lasing in this system is unidirectional, that is the phonon lasing takes place when the coupled system is driven in one direction but not the other.
Our work establishes the potential of spinning optomechanical devices for low-power mechanical isolation and unidirectional amplification.
This provides a new route, well within the reach of current experimental abilities, to operate cavity optomechanical devices for such a wide range of applications as directional phonon switches, invisible sound sensing, and topological or chiral acoustics.
\end{abstract}


\maketitle

\section{Introduction}

Cavity optomechanics (COM) \cite{Meystre2013,Metcalfe2014,Aspelmeyer2014} is playing an increasingly important role in making and steering on-chip devices, such as long-lived quantum memory \cite{Bagheri2011}, transducers \cite{Bagci2014,Tian2015,Lecocq2016}, motion sensing \cite{Arcizet2006,Srinivasan2011,yang2016}, and phonon lasing \cite{Vahala2009,Kent2010,Vahala2010,Mahboob2012,Kemiktarak2014,Painter2015,Xiao2017,Yang2018}. Phonon lasing, or coherent mechanical amplification, exhibits similar properties as those of an optical laser, such as threshold, gain saturation, and linewidth narrowing in the lasing regime \cite{Wu2013,Jing2014,wang2014,Lv2017}, as demonstrated in experiments with trapped ions, nano-beams, superlattices, resonators, or electromechanical devices \cite{Vahala2009,Kent2010,Vahala2010,Mahboob2012,Kemiktarak2014,Painter2015,Xiao2017,Yang2018}. It provides coherent acoustic sources to drive phononic devices for practical applications in e.g., audio filtering, acoustic imaging, or topological sound control \cite{imaging,filter,comb,topo,phononics}.
COM-based ultralow-threshold phonon lasers, featuring a continuously tunable gain spectrum to selectively amplify
phonon modes, from radio frequency to microwave rates \cite{Vahala2010,Xiao2017,Yang2018}, provide a particularly attractive setting to explore quantum acoustic effects, such as two-mode mechanical correlations \cite{Kemiktarak2014} or phononic sub-Poissonian distributions \cite{Painter2015}.

In parallel, nonreciprocal optics \cite{Potton2014,Zoller2017} has emerged as an indispensable tool for such a wide range of applications as invisibility cloaking, noise-free sensing, directional lasing, or one-way optical communications \cite{Lin2011,Feng2011,Scheucher2016,Miri2017,Sounas2017,Bahari2017}. Directional transmission of light has been achieved by using optical nonlinearities or dynamically-modulated media \cite{bi2011,chang2014,shi2015,yang2014,Shen2016,Verhagen2016,Hua2016,Painter2017}. As a crucial element in signal readout and information processing, directional optical amplifiers (with minimal noises from the output port) have also been proposed and studied, in microwave circuits \cite{qubit1,qubit2} or a non-Hermitian time-Floquet device \cite{Fleury2018}. In a recent experiment, a reconfigurable optical device was demonstrated \cite{Dong2018}, having switchable functions as either a circulator or a directional amplifier \cite{Clerk-prx,pr-app}.
Directional amplification of microwave signals has also been experimentally demonstrated in a multi-mode COM system  \cite{Sillanpaa}.
These abilities, allowing directional transmission and amplifications of optical signals \cite{Manipatruni2009,Bino2018,Yong2018,Bing2018,fan2018}, are fundamental for the emerging fields of chiral quantum optics and topological photonics \cite{Metelmann2017}.

As in optical systems \cite{qubit1,qubit2,Fleury2018,Dong2018,Clerk-prx,pr-app}, directional emissions and amplifications of phonons are particularly important in mechanical engineering \cite{li2011,Kim2015,Fleury2016,Alu2017,Poshakinskiy2017,Kim2016,Xu2016,Xu2017,thermal diode,Wang2018}, such as acoustic sensing or computing \cite{imaging,filter,comb,topo}. Here we propose a strategy to achieve a \textit{nonreciprocal mechanical amplifier} by coupling a COM resonator to a purely optical spinning resonator. We show that by exploiting the optical Sagnac effect \cite{Post1967,Chow1985,Ciminelli2017}, both the mechanical gain and the phonon-lasing threshold can be significantly altered. In particular, by driving the COM resonator from the left or the right side, coherent emission of phonons is enhanced or completely suppressed, enabling a highly-tunable nonreciprocal phonon laser. This provides a key element for applications of COM devices in e.g., chiral quantum acoustics or topological phononics \cite{comb,topo}.

\begin{figure}[ht]
\centering
\includegraphics[width=3.4in]{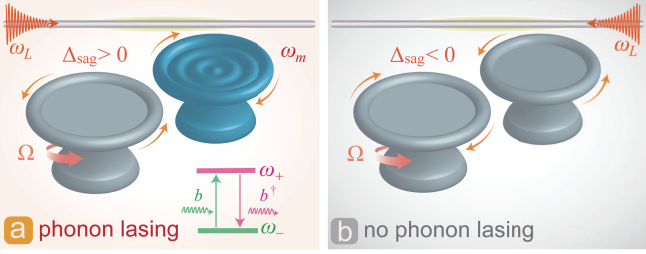}
\caption{(a) Schematic illustration of a nonreciprocal phonon laser composed of a spinning optical resonator coupled to a COM resonator. For the CCW rotation, driving the system from the left (with $\Delta_\mathrm{sag}>0)$ enhances the mechanical gain. The inset shows the equivalent two-level system model for phonon lasing where the transitions between supermodes are mediated by phonons. (b) Driving the system from the right (with $\Delta_\mathrm{sag}<0$) suppresses or even completely blocks the phonon-lasing process.}
\label{fig1}
\end{figure}

We note that very recently, by spinning a whispering-gallery-mode (WGM) optical resonator, nonreciprocal transmission with $99.6\%$ rectification was observed for photons, without any magnetic field or optical nonlinearity \cite{Carmon}. By spinning a shape-deformed resonator, purely-optical effects such as mode coupling \cite{Ge2014} and broken chiral symmetry \cite{Sarma2015} have also been revealed. Such spinning devices also can be useful in nano-particle sensing \cite{Jing2018} or single-photon control \cite{Huang2018}. For sound, excellent isolation was demonstrated even earlier by using a circulating fluid \cite{Fleury2014}. In recent experiments, chiral mechanical cooling was also realized via nonreciprocal Brillouin scattering \cite{Kim2015,Kim2016,Xu2016,Xu2017}.
Here, as another necessary element, we show that nonreciprocal \textit{amplifications} of phonons can be achieved by utilizing optical Sagnac effect induced by a spinning resonator \cite{Carmon}. This opens up a new route to operate nonreciprocal COM devices for applications in e.g., backaction-immune force sensing \cite{Vahala2009} and chiral acoustic engineering \cite{Zhang2018,Kuzyk}.

\section{Model and solutions}

We consider two coupled resonators, one of which is purely optical and the other supports a mechanical breathing mode (frequency $\omega_m$ and effective mass $m$) when pumped by light at frequency $\omega_L$ and amplitude $\varepsilon_L$.
Evanescent coupling between these resonators exists in the $1550\,\mathrm{nm}$ band, and the laser is coupled into and out of the resonator via a waveguide (see Fig.\,\ref{fig1}). The two resonators share the same resonance frequency $\omega_c$ for the stationary case. The decay rates of the COM resonator and the optical resonator are $\gamma_1$ and $\gamma_2$, respectively, which are related to the optical quality factors $Q_1$ and $Q_2$, i.e., $\gamma_{1,2}=\omega_c/Q_{1,2}$. This compound system, utilized to observe phonon lasing experimentally \cite{Vahala2010,Xiao2017}, can be further tuned by spinning the optical resonator with speed $\Omega$, due to the $\Omega$-dependent optical Sagnac effect. An immediate consequence of this effect is that for the counter-clockwise (CCW) rotation denoted by $\Omega>0$, \textit{phonon lasing can be enhanced or blocked} by driving the system from the left or the right (or equivalently, for the clockwise (CW) rotation denoted by $\Omega<0$, the phonon lasing is enhanced by driving the system from the right and it is blocked when the drive is from the left). This, as aforementioned, indicates a \textit{highly-tunable nonreciprocal phonon laser} by flexibly tuning the rotation speed and drive direction.

In a spinning resonator, the CW or CCW optical mode experiences different refractive indices \cite{Carmon}, i.e., $n_\pm=n[1\pm r_2\Omega(n^{-2}-1)/c]$, where $n$ and $r_2$ denote, respectively, the refractive index and the radius of the resonator, and $c$ is the speed of light in vacuum. As a result, the frequencies of the CW and CCW modes of the resonator experience Sagnac-Fizeau shifts \cite{Malykin,jing2017}. For light propagating in the same, or opposite, direction of the spinning resonator, we have  $$\omega_c\rightarrow\omega_c\mp|\Delta_\mathrm{sag}|,$$ with
\begin{align}
\Delta_\mathrm{sag}=\pm\Omega\frac{nr_2\omega_c}{c}\left(1-\frac{1}{n^2}-\frac{\lambda}{n}\frac{dn}{d\lambda}\right),
\label{eq:sagnac}
\end{align}
where $\lambda=c/\omega_c$ is the optical wavelength. The dispersion term $(\lambda/n)(dn/d\lambda)$ denotes the Lorentz correction to the Fresnel-Fizeau drag coefficient, characterizing the relativistic origin of the Sagnac effect \cite{Malykin}. This term, relatively small in typical materials, is confirmed to be safely ignored in the recent experiment \cite{Carmon}. In such a device, light is dragged by the spinning resonator, leading to nonreciprocal transmissions for optical counter-propagating modes (see Ref.\,\cite{Carmon} for more details). Below we show that for our COM system, this leads to distinct changes of the radiation pressure on the mechanical mode, hence resulting in a \textit{nonreciprocal phonon laser}.

As shown in Fig.\,\ref{fig1}, spinning the resonator along the CCW direction and driving the device from the left or the right, induces an optical red or blue shift, i.e., $\Delta_\mathrm{sag}>0$ or $\Delta_\mathrm{sag}<0$. The Hamiltonian of the system, in a frame rotating at the frequency $\omega_L$, can be written as ($\hbar=1$)
\begin{align}
H&=H_0+H_{\mathrm{int}}+H_{\mathrm{dr}},\nonumber\\
H_0&=-\Delta_L a_1^\dag a_1-(\Delta_L+\Delta_\mathrm{sag})a_2^\dag a_2+\omega_mb^\dag b,\nonumber\\
H_{\mathrm{int}}&=-\zeta x_0a_1^\dag a_1(b^\dag+b)+J(a^\dag_1a_2+a^\dag_2a_1),\nonumber\\
H_{\mathrm{dr}}&=i(\varepsilon_La_1^\dag-\varepsilon_L^*a_1).
\label{eq:Hamiltonian}
\end{align}
$H_0$ is the free Hamiltonian where the first and second terms describe the optical modes in the COM resonator and the spinning resonator, respectively; and the third term denotes the energy of the mechanical mode. $H_\mathrm{int}$ is the interaction Hamiltonian where the first term describes the coupling between the optical and mechanical mode in the optomechanical resonator; and the second term describes the coupling between the optical modes of the resonators. Note that $\Delta_L=\omega_L-\omega_c$ is the detuning between the drive laser and the resonance frequency of the resonator, $a_1$ and $a_2$ denote the optical annihilation operators of the resonators (coupled with the optical strength $J$), $x_0=\sqrt{\hbar/2m\omega_m}$, $b$ is the mechanical annihilation operator, $\zeta=\omega_c/r_1$ denotes the COM coupling strength, and $r_1$ is the radius of the COM resonator. Finally, $H_{\mathrm{dr}}$ denotes the drive which is fed into the coupled resonator system through the waveguide (see Fig.\,\ref{fig1}), with the driving amplitude
$$\varepsilon_L=\sqrt{2\gamma_1P_\mathrm{in}/\hbar\omega_L},$$
and the input power $P_\mathrm{in}$.

The Heisenberg equations of motion of this compound system are then written as
\begin{align}
\dot{a}_1&=(i\Delta_L-\gamma_1)a_1+i\zeta x_0(b+b^\dag)a_1-iJa_2+\varepsilon_L,\nonumber\\
\dot{a}_2&=[i(\Delta_L+\Delta_{\mathrm{sag}})-\gamma_2]a_2-iJa_1,\nonumber\\
\dot{b}&=-(i\omega_m+\gamma_m)b+i\zeta x_0a^\dag_1a_1.
\label{eq:steady}
\end{align}
Here, $\gamma_m$ is the damping rate of the mechanical mode. We remark that as already confirmed in the experiment on optomechanical phonon laser \cite{Vahala2010}, for a strong pump field, the quantum noise terms can be safely ignored, if one only concerns about the mean-number behaviors (i.e., the threshold feature of the mechanical gain or the phonon amplifications). Setting all the derivatives of Eq.\,(\ref{eq:steady}) as zero, the steady-state solutions of the system can be readily derived as
\begin{align}
a_{1,s}&=\frac{\varepsilon_L}{\gamma_1 -(i\Delta_L+\zeta x_s)+J^2/[\gamma_2-i(\Delta_L\pm|\Delta_{\mathrm{sag}}|)]},\nonumber\\
a_{2,s}&=\frac{Ja_{1,s}}{\Delta_L\pm|\Delta_{\mathrm{sag}}|+i\gamma_2},~~~b_s=\frac{\zeta x_0|a_{1,s}|^2}{\omega_m-i\gamma_m},
\end{align}
where $x_s=x_0(b_s+b^\ast_s)$ is the steady-state mechanical displacement. Combining these expressions gives the balance equation of the radiation and spring forces
$$
m(\omega_m^2+\gamma_m^2)x_s=\hbar\zeta|a_{1,s}|^2.
$$
The displacement $x_s$ is determined from the optical density $|a_{1,s}|^2$ inside the COM resonator, which clearly depends on $\Delta_\mathrm{sag}$ (see also Fig.\,\ref{fig2}, e.g., both $|a_{1,s}|^2$ and $x_s$ become significantly different for $\Delta_\mathrm{sag}>0$ or $\Delta_\mathrm{sag}<0$). Also the ratio $\eta$ of the steady-sate mechanical displacement $x_s$ for spinning and no spinning the resonator is given by
\begin{align}
\eta_{>,<}\equiv\frac{x_s(\Delta_{\mathrm{sag}}>0,\,<0)}{x_s(\Omega=0)}.\nonumber
\end{align}

In close analogy to an optical laser, a coherent emission of phonons can be achieved with compound resonators through inversion of the two optical supermodes \cite{Vahala2010,Xiao2017}. This leads to a \textit{phonon laser at the breathing mode} with frequency $\omega_m$, above the threshold power $P_\mathrm{th}\sim 7\,\mu \mathrm{W}$, according to Grudinin ${et~al.}$ \cite{Vahala2010}.

By using the supermode operators $a_{\pm}=(a_1\pm a_2)/\sqrt{2}$, $H_0$ and $H_{\mathrm{dr}}$ in Eq.\,(\ref{eq:Hamiltonian}) can be written as
\begin{align}
\mathcal{H}_0&=\omega_+a^\dag_+a_++\omega_-a^\dag_-a_-+\omega_mb^\dag b,\nonumber\\
\mathcal{H}_{\mathrm{dr}}&=\frac{i}{\sqrt{2}}[\varepsilon_L(a^\dag_++a^\dag_-)-\mathrm{H.c.}],
\end{align}
with the supermode frequencies $$\omega_{\pm}=-\Delta_L-\frac{1}{2}\Delta_\mathrm{\mathrm{sag}}\pm J.$$
Under the rotating-wave approximation \cite{Vahala2010,Jing2014}, the interaction Hamiltonian can be written  as
\begin{align}
\mathcal{H}_{\mathrm{int}}=&-\frac{\zeta x_0}{2}(a^{\dagger}_{+}a_{-}b+b^{\dagger}a^{\dagger}_{-}a_{+})\nonumber\\
&+\frac{\Delta_{\mathrm{sag}}}{2}(a^{\dagger}_{+}a_{-}+a^{\dagger}_{-}a_{+}).
\label{eq:int}
\end{align}
Besides the first term which describes the absorption and emission of phonons (as in a conventional COM system) \cite{Vahala2010}, $\mathcal{H}_{\mathrm{int}}$ in Eq.\,(\ref{eq:int}) includes an additional $\Omega$-dependent term which implies that the \textit{coupling} between the optical supermodes depends on the Sagnac effect. The second term in Eq.\,(\ref{eq:int}) is the reason for the striking modifications in the phonon-lasing process, which is very different from the ordinary cases without the coupling of supermodes \cite{Vahala2010}.
We note that in general, the supermode operators $a_\pm = (a_1\pm a_2)/\sqrt{2}$ are defined for coupled cavities sharing the same resonant frequency. These operators can still be used here due to the fact that the Sagnac shift in our system is much smaller than the optical detuning and the optical coupling rate.
It is possible to introduce another transformation to diagonalize the two-mode system, such as that in a recent work on phonon laser \cite{Yang2018}. We have confirmed that, since the Sagnac shift is $\Delta_\mathrm{sag}/\omega_m\simeq 0.1$ for $\Omega= 6\,$ kHz, i.e., much smaller than $\Delta_L$ and $J$, this transformation can be safely reduced to the above operators as we used.

In the supermode picture, we can define the \textit{ladder} operator and \textit{population inversion} operator of the optical supermodes as \cite{Vahala2010} $$p=a^\dag_- a_+,~~~\delta n=a^\dag_+ a_+-a^\dag_-a_-,$$ respectively. The equations of motion of the system then become
\begin{align}
\dot{b}=&-(\gamma_m+i\omega_m)b+\frac{i\zeta x_0}{2}p,\nonumber\\
\dot{p}=&-2(\gamma +iJ)p+\frac{i}{2}(\Delta_{\mathrm{sag}}-\zeta x_0b)\delta n \nonumber\\
&+\frac{1}{\sqrt{2}}(\varepsilon_L^*a_{+}+\varepsilon_La^{\dagger}_{-}),
\end{align}
with $\gamma=(\gamma_1+\gamma_2)/2$. By using the standard procedures (see Appendix B for more details), we can easily obtain the \textit{mechanical gain}, i.e., $G=G_0+\mathcal{G}$, where
\begin{align}
G_0=\frac{(\zeta x_0)^2\gamma\delta n}{2(2J-\omega_m)^2+8\gamma^2},
\label{eq:gain}
\end{align}
and
\begin{align}
\mathcal{G}=\frac{|\varepsilon_L|^2(\zeta x_0)^2(\omega_m-2J)(\Delta_\mathrm{sag}+2\Delta_L)\gamma}{4\left[ \beta^2+(2\Delta_L+\Delta_\mathrm{sag})^2\gamma^2\right]\left[(2J-\omega_m)^2+4\gamma^2\right]},
\end{align}
with
\begin{align}
\beta\simeq J^2+\gamma^2-\Delta_L^2+\frac{(\zeta x_0)^2n_b}{4}-\Delta_\mathrm{sag}\Delta_L,
\end{align}
in consideration of $\Delta_\mathrm{sag}\ll\Delta_L,J$ and $\zeta x_0/\Delta_L\ll1$. The \textit{population inversion} $\delta n$ can also be derived as
\begin{align}
\delta n
\simeq&\frac{2J|\varepsilon_L|^2}{\beta_0^2+4\gamma^2\Delta_L^2}\left(\Delta_L+\Delta_\mathrm{sag}\right),
\label{eq:inversion}
\end{align}
with $\beta_0=\beta(\Delta_\mathrm{sag}=0)$.
We have confirmed that the condition $\zeta x_0/\Delta_L\ll 1$ is valid for the range of parameters used in this work. However, in the numerical simulations, we have \textit{not} used this approximation and thus the presented results are valid for the general case. Different from the conventional phonon-laser system where both resonators are stationary \cite{Vahala2010}, besides the term $G_0$, we also have a new term $\mathcal{G}$ that depends on both $\Delta_\mathrm{sag}$ and $\Delta_L$. This indicates that by tuning $\Omega$ and $\Delta_L$ together, the \textit{mechanical gain $G$ could be made very different} for $\Delta_{\mathrm{sag}}>0$ or $\Delta_{\mathrm{sag}}<0$.

We note that the non-negative mechanical gain $G$ decreases the effective damping rate of the mechanical mode $\gamma_\mathrm{eff}=\gamma_m-G$. Initially, this leads to heating of the mechanical oscillator, and parametric instabilities can occur for $\gamma_\mathrm{eff}<0$. In this situation, an initial fluctuation of the mechanical displacement can grow exponentially until the oscillation amplitude is saturated due to the nonlinear effects, which results in a steady-state regime with a fixed oscillation amplitude (i.e., the phonon-lasing regime) \cite{Vahala2010,Aspelmeyer2014}. In practice, the in-phase and quadrature components of the mechanical motion mode, as well as its power spectral density, can be experimentally measured, from which a transition from a thermal state below threshold to a coherent state above threshold can be demonstrated, as the linear gain is turned on and allowed to increase until the phonon laser reaches the steady state \cite{Painter2015}.

\section{numerical results and discussions}

\begin{figure}[ht]
\centering
\includegraphics[width=3in]{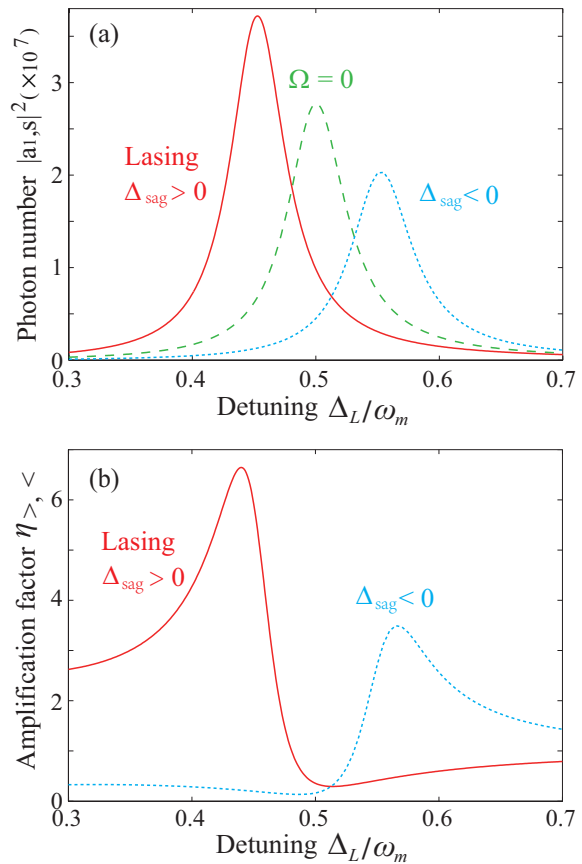}
\caption{(a) The steady-state photon number $|a_{1,s}|^2$ as a function of the optical detuning $\Delta_L$. The dependence of $|a_{1,s}|^2$ on the spinning speed $\Omega$ is clearly seen. (b) The mechanical displacement amplification factor $\eta$ versus $\Delta_L$. Parameters are chosen as $\Omega=6$\,kHz, $J/\omega_m=0.5$, and $P_\mathrm{in}=10\,\mu$W.}
\label{fig2}
\end{figure}

Figure\,\ref{fig2}(a) shows the steady-state populations of intracavity photons as a function of the optical detuning. As in relevant experiments \cite{Vahala2010,Carmon,Righini2011}, the parameter values are taken as: $n=1.48$, $r_1=34.5\,\mu$m, $Q_1=9.7\times10^7$, $r_2=4.75\,$mm, $Q_2=3\times10^7$, $m=50$\,ng, $\gamma_m=0.24$\,MHz, $\omega_m=2\pi\times23.4$\,MHz, $\Omega=6\,$kHz, and thus $\Delta_\mathrm{sag}/\omega_m\sim 0.1$. It is seen that spinning the resonator \textit{increases} the intracavity photon number $|a_{1,s}|^2$ when $\Delta_{\mathrm{sag}}>0$ or \textit{decreases} it when $\Delta_{\mathrm{sag}}<0$, compared to the stationary resonator case ($\Omega=0$). This change in the intracavity photon number then modifies the radiation pressure. Thus, we \textit{can tune (increase or decrease) the strength of optomechanical interactions effectively by tuning the speed and direction of the rotation of the resonator}. Intuitively, this direction-dependent feature for the intracavity photon number (and also the resulting radiation pressure) can be well understood by the motion-induced different refractive indices for the counter-propagating modes, as demonstrated very recently in a spinning resonator \cite{Carmon} (see also similar phenomena in a moving optical lattice \cite{Ramezani2018} or in an acoustic device with a circulating fluid \cite{Fleury2014}).

\begin{figure}[h]
	\centering
	\includegraphics[width=3in]{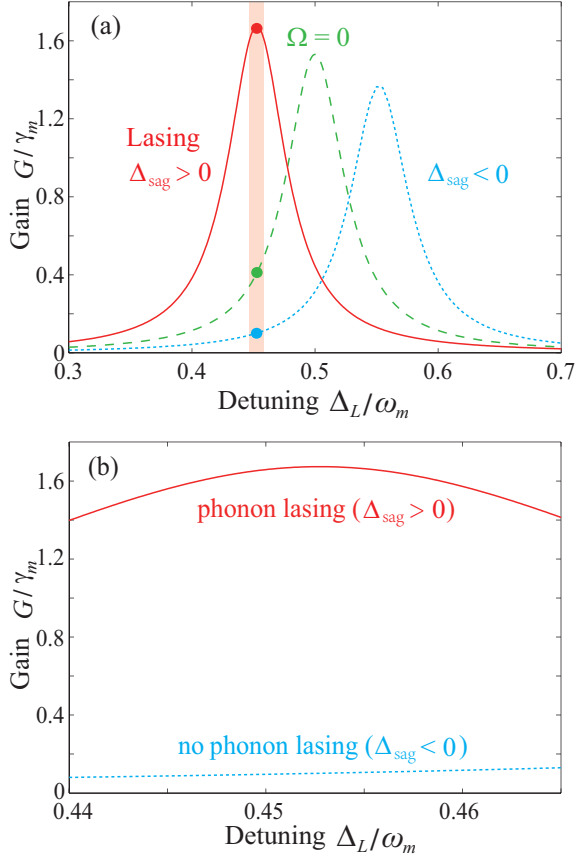}
	\caption{The scaled mechanical gain $G/\gamma_m$ as a function of the optical detuning $\Delta_L$. The phonon lasing can be generated (red curve) or prohibited (blue curve) for different rotation directions, indicating that the acoustic nonreciprocity can be achieved, as shown in (b). The thick points in the yellow vertical band correspond to two different driving directions at $\Delta_L/\omega_m=0.45$. The other parameters are the same as those in Fig.\,\ref{fig2}.}
	\label{fig3}
\end{figure}

Figure\,\ref{fig2}(b) shows the mechanical displacement amplification factor $\eta$. Note that $x_s$ is enhanced in the red detuning regime for $\Delta_{\mathrm{sag}}>0$, or the blue detuning regime for $\Delta_{\mathrm{sag}}<0$, which is due to the enhanced COM interaction. The amplified displacement indicates an enhancement of the phonon generation.

\begin{figure}[ht]
\centering
\includegraphics[width=3in]{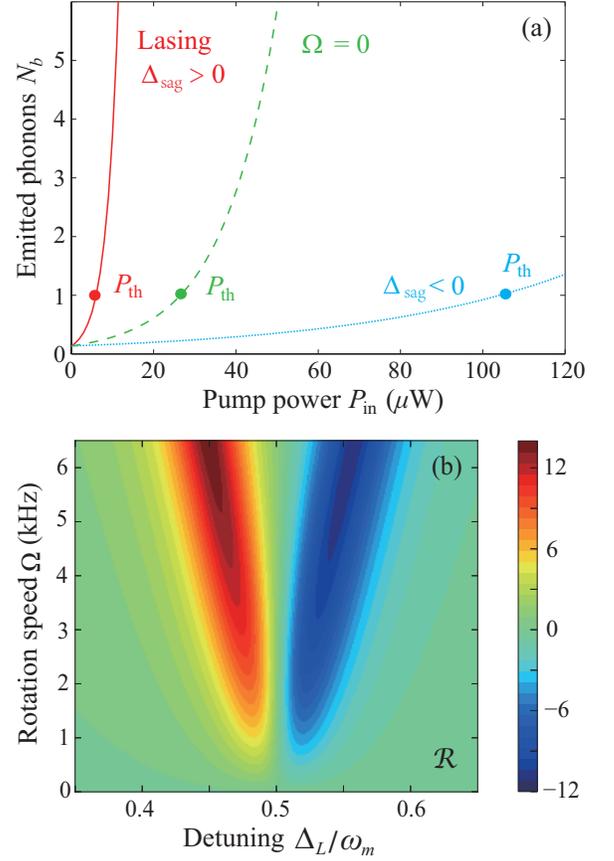}
\caption{(a) The stimulated emitted phonon number $N_b$ as a function of the pump power $P_{\mathrm{in}}$. The thick points correspond to the threshold power $P_\mathrm{th}$, which is determined by the threshold condition $G=\gamma_m$. (b) Dependence of the isolation parameter $\mathcal{R}$ on the optical detuning $\Delta_L$ and the rotation speed $\Omega$. The isolation parameter $\mathcal{R}$ can be maximized as $\mathcal{R}\sim10\,$dB for $\Delta_L/\omega_m=0.45$ and $\Omega=6\,$kHz [red point in Fig.\,\ref{fig3}(a)]. The parameters are $J/\omega_m=0.5$ in (a,b), and $P_\mathrm{in}=10\,\mu$W in (b).}
\label{fig4}
\end{figure}

We show in Fig.\,\ref{fig3} the mechanical gain $G$ as a function of the optical detuning $\Delta_L$, for different values of $\Omega$. In the stationary resonator case $(\Omega=0)$, the peak position of $G$ is always the same regardless of the direction of the driving light: we have $G>\gamma_m$ around $\Delta_L/\omega_m\sim 0.5$, corresponding to a conventional phonon laser in the blue-detuning regime \cite{Vahala2010}. In contrast, spinning the resonator leads to a $red$ or $blue$ shift also for the mechanical gain $G$, with $\Delta_{\mathrm{sag}}>0$ or $\Delta_{\mathrm{sag}}<0$, respectively. Due to these shifts, by tuning $\Delta_L$ (e.g., $\Delta_L/\omega_m\sim 0.45$ in the specific example of Fig.\,\ref{fig3}), the mechanical gain can be enhanced for $\Delta_{\mathrm{sag}}>0$, while significantly suppressed (i.e., $G<\gamma_m$) for $\Delta_{\mathrm{sag}}<0$.

The underlying physics can be explained as follows. Spinning the resonator results in opposite shifts of the counter-propagating WGMs, leading to nonreciprocal light transmission \cite{Carmon}. For the $\Delta_{\mathrm{sag}}>0$ case, the driving light is in the resonators, inducing an \textit{enhanced radiation pressure}, which corresponds to an enhanced population inversion [see Eq.\,(\ref{eq:inversion})]. As a result, the mechanical gain is enhanced.
For the $\Delta_{\mathrm{sag}}<0$ case, on the other hand, the driving light is transmitted out of the resonators, inducing a weakened radiation pressure, so that the mechanical gain is nearly zero, i.e., no phonon lasing.
Thus, our system provides a new route to control the behavior of phonon lasing.

Once the mechanical gain is obtained, the \textit{stimulated emitted phonon number} $N_b$ can be calculated, i.e.,
\begin{equation}
N_b=\mathrm{exp}[2(G-\gamma_m)/\gamma_m],
\end{equation}
which characterizes the performance of the phonon laser. Figure\,\ref{fig4}(a) shows $N_b$ with $\Delta_L/\omega_m=0.45$ and $\Omega=6$\,kHz [corresponding to the maximal value of the mechanical gain in Fig.\,\ref{fig3}(a)]. From the threshold condition for the phonon lasing $N_b=1$, i.e., $G=\gamma_m$, we can easily derive the threshold pump power \cite{Vahala2010}. For $J/\omega_m=0.5$, we substitute Eqs.\,(\ref{eq:gain}) and (\ref{eq:inversion}) into the threshold condition, and then obtain
\begin{align}
P_\mathrm{th}\approx\frac{2\hbar\gamma\gamma_m\omega_c [M+\gamma^2(2\Delta_L+\Delta_\mathrm{sag})^2]}{\gamma_1 J(\zeta x_0)^2(\Delta_L+\Delta_\mathrm{sag})},
\end{align}
with $$M=(J^2+\gamma^2-\Delta^2_L-\Delta_\mathrm{sag}\Delta_L)^2,$$
in which we have used $|b_s|^2\ll 1$ at the threshold. We can see that the Sagnac effect has a significant impact on the threshold.
For $\Delta_{\mathrm{sag}}>0$, the threshold power can be reduced to $6.03\,\mu$W, which is attributed to the enhancement of the mechanical gain. It is obvious that more phonons can be generated with larger pump powers. For $\Delta_{\mathrm{sag}}<0$, the mechanical gain is blocked at $\Delta_L/\omega_m=0.45$, so that a larger input power is needed to achieve the phonon laser.
We note that the threshold power is up to about $26.6\,\mu$W for a stationary system ($\Omega=0$). However, at the resonance point ($\Delta_L/\omega_m=0.5$), the threshold for phonon lasing is $6.53\,\mu$W, which approaches the threshold of about $7\,\mu$W reported in experiments \cite{Vahala2010}.

An \textit{optimal spinning speed} should be chosen to obtain a considerable phonon number. In order to clearly see the effect of the spinning speed on the stimulated emitted phonon number, we introduce the isolation parameter \cite{Hua2016,Alu2017}
\begin{align}
\mathcal{R}=10\log_{10}\frac{N_b(\Omega>0)}{N_b(\Omega<0)}.
\end{align}
\textit{Acoustic nonreciprocity} can be achieved for $\mathcal{R}\neq 0$ or  $$N_b(\Delta_{\mathrm{sag}}>0)\neq N_b(\Delta_{\mathrm{sag}}<0),$$ indicating that the spinning COM system is driven from two different directions.

Figure\,\ref{fig4}(b) shows $\mathcal{R}$ versus the optical detuning and spinning speed. Nonreciprocity emerges for the two detuning regions around $\Delta_L/\omega_m\sim 0.45$ and $\Delta_L/\omega_m\sim 0.55$. For $\Delta_L/\omega_m=0.5$, we have $\mathcal{R}\sim 0$ implying a reciprocal system.
The nonreciprocity becomes obvious for the spinning resonator, which is an inevitable result from the difference between $\delta n(\Delta_{\mathrm{sag}}>0)$ and $\delta n(\Delta_{\mathrm{sag}}<0)$. Phonon lasing is favorable to be generated in the $\Delta_{\mathrm{sag}}>0$ regime and is always suppressed for $\Delta_{\mathrm{sag}}<0$. This brings about the convenience of turning on or off the phonon lasing just by changing the driving direction.
We note that optical nonreciprocity has been demonstrated in a pure optical system by spinning the resonator \cite{Carmon}.
In our spinning COM device, nonreciprocal phonon lasing can be realized due to the optical Sagnac effect, which can change the radiation pressure in the COM devices.

\section{summary}

In summary, we have studied theoretically the role of rotation in engineering a nonreciprocal phonon laser. We show that in our system, consisting of a COM resonator coupled to a spinning optical resonator \cite{Vahala2010,Carmon}, the optical Sagnac effect strongly modifies not only the intracavity optical intensities but also the mechanical gain and the phonon-lasing threshold. As a result, the threshold pump power can be reduced or raised, depending on whether the drive is input in the same or opposite direction of the spinning resonator. Our results, i.e., controlling the behavior of a phonon laser by using a spinning resonator, shed new light on engineering COM or other acoustic devices, such as COM transducers or motion sensors. In our future works, we will further study e.g. purely quantum correlations of emitted phonons, in which quantum noise terms should be included, or a nonreciprocal phonon laser operating at an exceptional point \cite{Yang2018}.

\bigskip

\section*{ACKNOWLEDGMENTS}
We thank S. K. \"Ozdemir at Pennsylvania State University for helpful discussions.
Y.J. and H.J. are supported by NSFC (11474087, 11774086). S.M. and T.C. are supported by the Israeli Centers of Research Excellence Circle of Light and the Israeli Science Foundation.
F.N. is supported by the MURI Center for Dynamic Magneto-Optics via the Air Force Office of Scientific Research (FA9550-14-1-0040), Army Research Office (Grant No. 73315PH), Asian Office of Aerospace Research and Development (Grant No. FA2386-18-1-4045), Japan Science and Technology Agency (the ImPACT program and CREST Grant No. JPMJCR1676), Japan Society for the Promotion of Science (JSPS-RFBR Grant No. 17-52-50023), RIKEN-AIST Challenge Research Fund, and the John Templeton Foundation.


\section*{APPENDIX A: DERIVATION OF THE HAMILTONIAN}

We consider two coupled WGM resonators, as shown in Fig.\,\ref{fig1}. One resonator supports a mechanical mode with frequency $\omega_m$ and effective mass $m$, which is pumped by a driving field at frequency $\omega_L$. The other resonator is purely optical, which can spin. The optical modes in the COM and optical resonators are denoted as $a_1$ and $a_2$, respectively.

The two cavities share the same resonance frequency (denoted as $\omega_c$) for the stationary case.
The resonance frequency of the spinning optical resonator can be shifted, as a result of the Sagnac effect. Therefore, the free Hamiltonian describing the optical and mechanical modes can be written as ($\hbar=1$)
\begin{align}
H^{\prime}_0=\omega_c a_1^\dag a_1+(\omega_c-\Delta_\mathrm{sag})a_2^\dag a_2+\omega_mb^\dag b,
\end{align}
where $b$ is the mechanical annihilation operator, and $\Delta_\mathrm{sag}$ is the frequency shift induced by the Sagnac effect.
For the resonator spinning along the CCW direction, $\Delta_\mathrm{sag}>0$ or $\Delta_\mathrm{sag}<0$ corresponds to the driving field from the left or the right.

In this system, we consider the coupling between the optical and mechanical mode in the COM resonator, and the evanescent coupling between the two resonators. The interaction Hamiltonian can be written as
\begin{align}
H^{\prime}_{\mathrm{int}}=-\zeta x_0a_1^\dag a_1(b^\dag+b)+J(a^\dag_1a_2+a^\dag_2a_1),
\end{align}
where $\zeta=\omega_c/r_1$ denotes the COM coupling strength, $J$ is the optical coupling strength, and $x_0=\sqrt{\hbar/2m\omega_m}$.
The driving field is fed into the COM resonator through the waveguide. Then the driving Hamiltonian reads
\begin{align}
H^{\prime}_{\mathrm{dr}}=i(\varepsilon_L e^{-i\omega_L t} a_1^\dag-\varepsilon^\ast_Le^{i\omega_L t} a_1),
\end{align}
where $\varepsilon_L=\sqrt{2\gamma_1P_\mathrm{in}/\hbar\omega_L}$ is the driving amplitude with the input power $P_\mathrm{in}$ and the optical loss rate $\gamma_{1}$.

The total Hamiltonian of the system can be written as $$H^{\prime}=H^{\prime}_0+H^{\prime}_{\mathrm{int}}+H^{\prime}_{\mathrm{dr}}.$$
By using the unitary transformation $$U=e^{-i\omega_L t(a_{1}^{\dagger} a_{1}+a_{2}^{\dagger}a_{2})},$$
the Hamiltonian $H^{\prime}$ can be transformed into the rotating frame, i.e.,
$$H=U^{\dagger}H^{\prime}U-iU^{\dagger}\frac{\partial U}{\partial t}.$$
Then we have
\begin{align}
H&=H_0+H_{\mathrm{int}}+H_{\mathrm{dr}},\nonumber\\
H_0&=-\Delta_L a_1^\dag a_1-(\Delta_L+\Delta_\mathrm{sag})a_2^\dag a_2+\omega_mb^\dag b,\nonumber\\
H_{\mathrm{int}}&=-\zeta x_0a_1^\dag a_1(b^\dag+b)+J(a^\dag_1a_2+a^\dag_2a_1),\nonumber\\
H_{\mathrm{dr}}&=i(\varepsilon_La_1^\dag-\varepsilon_L^*a_1),
\label{Aeq:Ha}
\end{align}
where $\Delta_L=\omega_L-\omega_c$ is the detuning of the driving field. This Hamiltonian sets the stage for our calculations of the mechanical gain and the threshold power.

We then introduce the supermode operators $a_{\pm}=(a_1\pm a_2)/\sqrt{2}$, which satisfy the commutation relations
\begin{align}
[a_{+}, a_{+}^\dag]=[a_{-}, a_{-}^\dag]=1,~
[a_{+}, a_{-}^\dag]=0.\nonumber
\end{align}
$H_0$ and $H_{\mathrm{dr}}$ in Eq.\,(\ref{Aeq:Ha}) can be written as
\begin{align}
\mathcal{H}_0&=\omega_+a^\dag_+a_++\omega_-a^\dag_-a_-+\omega_mb^\dag b,\nonumber\\
\mathcal{H}_{\mathrm{dr}}&=\frac{i}{\sqrt{2}}[\varepsilon_L(a^\dag_++a^\dag_-)-\mathrm{H.c.}],
\end{align}
with the frequencies $\omega_{\pm}=-\Delta_L-\frac{1}{2}\Delta_\mathrm{\mathrm{sag}}\pm J,$
and $H_{\mathrm{int}}$ can be transformed to
\begin{align}
\mathcal{H}_{\mathrm{int}}
=&\mathcal{H}_{\mathrm{int}}^{0}+\mathcal{H}_{\mathrm{int}}^1\nonumber\\
=&-\frac{\zeta
x_0}{2}[(a_{+}^{\dagger}a_++a_{-}^{\dagger}a_-)+(a_+^{\dagger}a_-+a_-^{\dagger}a_+)](b^{\dagger}+b)\nonumber\\
&+\frac{\Delta_\mathrm{sag}}{2}(a^\dag_+a_-+a^\dag_-a_+).
\end{align}
In the rotating frame with respect to $\mathcal{H}_0$, we have
\begin{align}
\mathcal{H}_{\mathrm{int}}^0=&-\frac{\zeta
x_0}{2}\Big[a_+^{\dagger}a_-b\mathrm{e}^{i(2J-\omega_m)t}+a_+a_-^{\dagger}b^{\dagger}\mathrm{e}^{-i(2J-\omega_m)t}\nonumber\\
&+a_+^{\dagger}a_-b^{\dagger}\mathrm{e}^{i(2J+\omega_m)t}+a_+a_-^{\dagger}b\mathrm{e}^{-i(2J+\omega_m)t}\Big]\nonumber\\
&+(a_+^{\dagger}a_++a_-^{\dagger}a_-)(b^{\dagger}\mathrm{e}^{i\omega_mt}+b\mathrm{e}^{-i\omega_mt})\Big].\nonumber
\end{align}
Under the rotating-wave approximation condition $$2J+\omega_m,\,\omega_m\gg|2J-\omega_m|,$$ the terms $a^\dag_{+}a_{-}b^\dag\mathrm{e}^{i(2J+\omega_m)t}$, $a_{+}a^\dag_{-}b\mathrm{e}^{-i(2J+\omega_m)t}$ and also $(a^\dag_{+}a_{+}+a^\dag_{-}a_{-})(b^\dag\mathrm{e}^{i\omega_mt}+b\mathrm{e}^{-i\omega_mt})$ can be omitted, in comparison with the near-resonance terms $a_+^{\dagger}a_-b\mathrm{e}^{i(2J-\omega_m)t}$ and $a_+a_-^{\dagger}b^{\dagger}\mathrm{e}^{-i(2J-\omega_m)t}$ \cite{Vahala2010}. Therefore, we have a simplified interaction Hamiltonian
\begin{align}
\mathcal{H}_{\mathrm{int}}=&-\frac{\zeta x_0}{2}(a^{\dagger}_{+}a_{-}b+b^{\dagger}a^{\dagger}_{-}a_{+})\nonumber\\
&+ \frac{\Delta_{\mathrm{sag}}}{2}(a^{\dagger}_{+}a_{-}+a^{\dagger}_{-}a_{+}).
\label{Aeq:int}
\end{align}

\section*{APPENDIX B: DERIVATION OF THE MECHANICAL GAIN}

In the supermode picture, the equations of motion of the system can be written as
\begin{align}
\dot{a}_+&=-(i\omega_++\gamma)a_++\frac{i}{2}(\zeta x_0b-\Delta_\mathrm{sag})a_-+\frac{\varepsilon_L}{\sqrt{2}},\nonumber\\
\dot{a}_-&=-(i\omega_-+\gamma)a_-+\frac{i}{2}(\zeta x_0b^\dag-\Delta_\mathrm{sag})a_++\frac{\varepsilon_L}{\sqrt{2}},\nonumber\\
\dot{b}&=-(i\omega_m+\gamma_m)b+\frac{i\zeta x_0}{2}a_+a^\dag_-.
\end{align}
We can define the ladder operator and population inversion operator as
$$p=a_{-}^\dag a_{+},~~~\delta n=a_{+}^\dag a_{+}-a_{-}^\dag a_{-},$$
respectively. The equations of the system then read
\begin{align}
\dot{b}=&-(\gamma_m+i\omega_m)b+\frac{i\zeta x_0}{2}p,\nonumber\\
\dot{p}=&-2(\gamma +iJ)p+\frac{i}{2}(\Delta_{\mathrm{sag}}-\zeta x_0b)\delta n \nonumber\\
&+\frac{1}{\sqrt{2}}(\varepsilon_L^*a_{+}+\varepsilon_La^{\dagger}_{-}).
\label{dynamics}
\end{align}

By setting the time derivatives of $a_{\pm}$ and $p$ as zero with $\gamma\gg\gamma_m$, we obtain the steady-state values of the system, i.e.,
\begin{align}
p&=\frac{\sqrt{2}(\varepsilon^*_La_++\varepsilon_La^\dagger_-)-i(\zeta x_0b-\Delta_\mathrm{sag})\delta n}{2i(2J-\omega_m)+4\gamma},\nonumber\\
a_+&=\frac{\varepsilon_L(2i\omega_-+2\gamma+i\zeta x_0b-i\Delta_\mathrm{sag})}{2\sqrt{2}[\beta-i(2\Delta_L+\Delta_\mathrm{sag})\gamma]},\nonumber\\
a_-&=\frac{\varepsilon_L(2i\omega_++2\gamma+i\zeta x_0b^\dag-i\Delta_\mathrm{sag})}{2\sqrt{2}[\beta-i(2\Delta_L+\Delta_\mathrm{sag})\gamma]},
\label{steaty}
\end{align}
with
\begin{align}
\beta=&\beta_0-\Delta_\mathrm{sag}\left[\Delta_L+\frac{\zeta x_0}{2}\mathrm{Re}(b)\right],\nonumber\\
\beta_0=&J^2+\gamma^2-\Delta_L^2+\frac{(\zeta x_0)^2n_b}{4},\nonumber
\end{align}
and the phonon number $n_b=b^\dag b$.
Substituting Eq.\,(\ref{steaty}) into the dynamical equation of $b$ in Eq.\,(\ref{dynamics}) results in
\begin{align}
\dot{b}&=(-i\omega_m-i\omega'+G-\gamma_m)b+D,
\end{align}
where
\begin{align}
\omega'=&\frac{(\zeta x_0)^2(2J-\omega_m)\delta n}{4(2J-\omega_m)^2+16\gamma^2}\nonumber\\
&+\frac{(\zeta x_0)^2|\varepsilon_L|^2\gamma^2(2\Delta_L+\Delta_\mathrm{sag})}{[2(2J-\omega_m)^2+8\gamma^2][\beta^2+(2\Delta_L+\Delta_\mathrm{sag})^2\gamma^2]},\nonumber\\
D=&\frac{\zeta x_0\Delta_\mathrm{sag}\delta n}{4i(2J-\omega_m)+8\gamma}\nonumber\\
&+\frac{i\zeta x_0\beta(\gamma-iJ)|\varepsilon_L|^2}{[2i(2J-\omega_m)+4\gamma][\beta^2+(2\Delta_L+\Delta_\mathrm{sag})^2\gamma^2]}\nonumber\\
&+\frac{i\zeta x_0\gamma|\varepsilon_L|^2(2\Delta_L+\Delta_\mathrm{sag})(\Delta_L+\Delta_\mathrm{sag})}{[2i(2J-\omega_m)+4\gamma][\beta^2+(2\Delta_L+\Delta_\mathrm{sag})^2\gamma^2]},\nonumber
\end{align}
and the mechanical gain is $G=G_0+\mathcal{G},$ with
\begin{align}
G_0&=\frac{(\zeta x_0)^2\gamma\delta n}{2(2J-\omega_m)^2+8\gamma^2},\\
\mathcal{G}&=\frac{|\varepsilon_L|^2(\zeta x_0)^2(\omega_m-2J)(\Delta_\mathrm{sag}+2\Delta_L)\gamma}{4\left[ \beta^2+(2\Delta_L+\Delta_\mathrm{sag})^2\gamma^2\right]\left[(2J-\omega_m)^2+4\gamma^2\right]},\nonumber
\end{align}
where $\delta n$ can be expressed as
\begin{align}
\delta n
=&\frac{|\varepsilon_L|^2\left[2J(\Delta_L+\Delta_{\mathrm{sag}})-\gamma\zeta x_0\mathrm{Im}(b)-J\zeta x_0\mathrm{Re}(b)\right]}{\beta^2+\gamma^2(2\Delta_L+\Delta_{\mathrm{sag}})^2}\nonumber.
\end{align}
In consideration of $\Delta_\mathrm{sag}\ll\Delta_L,J$ and $\zeta x_0/\Delta_L\ll1$, we have
\begin{align}
\delta n\simeq&\frac{|\varepsilon_L|^2\left[2J(\Delta_L+\Delta_{\mathrm{sag}})-\gamma\zeta x_0\mathrm{Im}(b)-J\zeta x_0\mathrm{Re}(b)\right]}{\beta_0^2+4\gamma^2\Delta_L(\Delta_L-\Delta_\mathrm{sag})+\beta_0\Delta_{\mathrm{sag}}[\Delta_L+\zeta x_0\mathrm{Re}(b)]}\nonumber\\
\simeq&|\varepsilon_L|^2\cdot\frac{2J\Delta_L-\gamma\zeta x_0\mathrm{Im}(b)-J\zeta x_0\mathrm{Re}(b)-2J\Delta_{\mathrm{sag}}}{\beta_0^2+4\gamma^2\Delta_L^2}\nonumber\\
&\cdot\left[1-\frac{\beta_0\Delta_{\mathrm{sag}}(2\Delta_L+\zeta x_0\mathrm{Re}(b))-8\gamma^2\Delta_L\Delta_{\mathrm{sag}}}{\beta_0^2+4\gamma^2\Delta_L^2}\right]\nonumber\\
\simeq&\frac{|\varepsilon_L|^2[2J\Delta_L-J\zeta x_0\mathrm{Re}(b)-\gamma\zeta x_0\mathrm{Im}(b)]}{\beta_0^2+4\gamma^2\Delta_L^2}\nonumber\\
&\cdot\left(1+\frac{2\Delta_{\mathrm{sag}}\beta_0 \Delta_L}{\beta_0^2+4\gamma^2\Delta_L^2}\right)
+\frac{2\Delta_\mathrm{sag}J|\varepsilon_L|^2}{\beta_0^2+4\gamma^2\Delta_L^2}\nonumber\\
\simeq&\frac{2J|\varepsilon_L|^2}{\beta_0^2+4\gamma^2\Delta_L^2}(\Delta_L+\Delta_{\mathrm{sag}}), \nonumber
\end{align}
in which we have used $\beta\simeq \beta_0-\Delta_\mathrm{sag}\Delta_L$.

\section*{APPENDIX C: EXPERIMENTAL FEASIBILITY OF THE SPINNING RESONATOR}

The resonator can be mounted on a turbine, which spins the resonator, as in a very recent experiment \cite{Carmon}. In this experiment, the resonator with the radius $r=4.75\,$mm can spin with the stability of its axis, reaching the rotation frequency 3\,kHz. In our calculations, the rotation speed is chosen according to this experiment \cite{Carmon}.
For example, the Sagnac shift is $\Delta_\mathrm{sag}=14.6\,$MHz for $\Omega=6\,$kHz, leading to $\Delta_\mathrm{sag}/\omega_m\sim 0.1$.

By positioning the resonator near a single-mode fiber, the light can be coupled into or out the resonator evanescently through the tapered region. In the device, aerodynamic processes lead to a stable resonator-fiber coupling, which can be explained as follows. A fast spinning resonator can drag air into the region between the cavity and the taper, forming a boundary layer of air. Due to the air pressure on the surface of the taper facing the resonator, the taper flies at a height above the resonator, which can be several nanometers. If some perturbation induces the taper rising higher than the stable equilibrium height, it floats back to its original position \cite{Carmon}. The self-adjustment of the taper separation from the spinning resonator enables critical coupling of light into the cavity, by which counter-circulating lights experience optical drags identical in size, but opposite in sign. This experiment also confirms that the taper did not touch or stick to the rotating resonator even if the taper is pushed towards it, which is in contrast to the situation for a stationary resonator (i.e., the taper can stick to the resonator through van der Waals forces and thus needs to be pulled back to break the connection). Other factors, including intermolecular forces, lubricant compressibility, tapered-fiber stiffness and wrap angle of the fiber, may affect the resonator-waveguide coupling. However, these factors are confirmed to be negligible in the experiment.
In our scheme, the spinning resonator is coupled with the stationary COM resonator, instead of the fiber, in which stationary coupling of the two resonators can also be achieved.

\end{document}